# Toward a terahertz-driven electron gun

W. Ronny Huang[1], Emilio A. Nanni[1], Koustuban Ravi[1], Kyung-Han Hong[1], Liang Jie Wong[1], Phillip D. Keathley[1], A. Fallahi[2], Luis E. Zapata[2], and Franz X. Kärtner[1,2,]*

[1]*Department of Electrical Engineering and Computer Science and Research Laboratory of Electronics, Massachusetts Institute of Technology, Cambridge, Massachusetts 02139, USA*

[2]*Center for Free-Electron Laser Science, DESY and Department of Physics, University of Hamburg, Hamburg 22607, Germany*

*Corresponding author. Email: kaertner@mit.edu.

**Abstract**

**Femtosecond electron bunches with keV energies and eV energy spread are needed by condensed matter physicists to resolve state transitions in carbon nanotubes [1], molecular structures [2], organic salts [3], and charge density wave materials [4]. These semirelativistic electron sources are not only of interest for ultrafast electron diffraction [5,6], but also for electron energy-loss spectroscopy [1,7] and as a seed for x-ray FELs [8]. Thus far, the output energy spread (hence pulse duration) of ultrafast electron guns has been limited by the achievable electric field at the surface of the emitter [6], which is 10 MV/m for DC guns [9] and 200 MV/m for RF guns [10]. A single-cycle THz electron gun provides a unique opportunity to not only achieve GV/m surface electric fields but also with relatively low THz pulse energies, since a single-cycle transform-limited waveform is the most efficient way to achieve intense electric fields. Here, electron bunches of 50 fC from a flat copper photocathode are accelerated from rest to tens of eV by a microjoule THz pulse with peak electric field of 72 MV/m at 1 kHz repetition rate. We show that scaling to the readily-available GV/m THz field regime [11,12] would translate to monoenergetic electron beams of ~100 keV.**

**Introduction**

The realization of a compact electron gun requires very intense electric fields in order to achieve electron bunches with high charge density, good beam quality, and large kinetic energy. Conventional DC guns reach their performance limit at 10 MV/m due to field emission breakdown [9], while RF guns operate at 50-110 MV/m due to plasma breakdown limitations [10]. Since the plasma breakdown threshold scales as the square root of the frequency [13], there has been great interest in the development of compact accelerators operating in higher-frequency regions of the electromagnetic spectrum [14,15]. Near-infrared (NIR) laser pulses are one option because of their high fields (multi-GV/m), accessibility, and relatively low cost sources. Various NIR guns based on evanescent waves [16] and accelerators based on dielectric structures [17,18] have been investigated. However, due to the short NIR wavelength, phase matching between the electromagnetic field and electron, space charge effects, limited charge density, and fabrication of accelerating structures present challenging hurdles. Laser plasma wakefield acceleration has achieved GeV electrons [19,20,21] using 100 TW - PW laser facilities at low repetition rates with percent-level energy spread and jitter. Accelerators employing radiation at THz frequencies hold promise because of their long wavelength and high field breakdown threshold, which can be as high as 27 GV/m at 0.47 THz [22] for common accelerating materials. Additionally, the THz sources can be pumped by compact sub-ps lasers which avoid the high voltage shielding limitations of DC guns and are less bulky and expensive than the klystrons which pump RF guns. Further, timing synchronization challenges common to accelerator pump-probe experiments are completely mitigated because the same laser source used to generate THz can be used as a probe. Finally, the advent of efficient THz generation techniques [23,24,25] has made THz based accelerators a realistic possibility and has motivated studies for using THz radiation in improving electron energy and beam properties in high-brightness linacs [26,27] and proton post-accelerators [28].

Thus far, keV electron acceleration of semi-relativistic electron bunches in a centimeter-long linac using single-cycle (SC) THz pulses has been demonstrated [29]. SC THz sources now routinely achieve multi-GV/m peak fields from only sub-mJ THz pulse energies [11,12], reducing energy consumption and Joule heating issues common to conventional RF guns. To achieve the highest fields in the initial acceleration stage, an electron gun driven by a SC THz pulse can be implemented. In this regard, very recently Li et al. and Wimmer et al. demonstrated THz acceleration (streaking) from excited atoms [30] and an isolated nanotip [31], respectively. While the basic physics of their work, like ours, demonstrates strong THz field-induced electron manipulation from rest, it is geared toward understanding low-charge (<0.1 fC) electron emission from atoms and nanostructures. Here, we aim to build a THz electron gun capable of accelerating femtocoulomb bunches to keV energies. Unlike the earlier experiments, we for the first time explore the effects of space charge and GV/m acceleration with THz fields. As a proof-of-concept, we demonstrate a mean acceleration of 18 eV on 50 fC electron bunches using a 6 µJ, SC THz field with peak field of 72 MV/m (ponderomotive energy of 28.5 eV). The experimentally obtained acceleration results are in good agreement with particle tracking simulations. To demonstrate the scalability of this approach, we show via simulation that a 2 GV/m THz pulse can achieve a mean energy of ~100 keV with a RMS energy spread of 1.3%.

**Results**

In the experiment, schematic and photograph shown in Figs. 1(a) and 1(b), an emit-and-accelerate scheme is employed on a flat copper photocathode, akin to the cathodes used in conventional accelerators (e.g. LCLS [32]) and well suited to high-electron-brightness applications due to its simplicity and robustness. Electrons are emitted by two-photon ionization at the photocathode using a 525 fs pulse at 515 nm (green) at 1 kHz repetition rate with a total charge per bunch of 50 fC. The photocurrent was measured at the photocathode. The emitted electrons were exposed to a p-polarized, SC THz pulse with a carrier frequency of 0.45 THz (Fig. 2(a)). The reflection from the photocathode increases the THz field twofold. The THz pulse delivered onto the copper is focused to a beam waist of 1.1 mm (Fig. 1(c)) and has a final energy of 6 µJ with a calculated peak electric field on the surface of 72 MV/m.

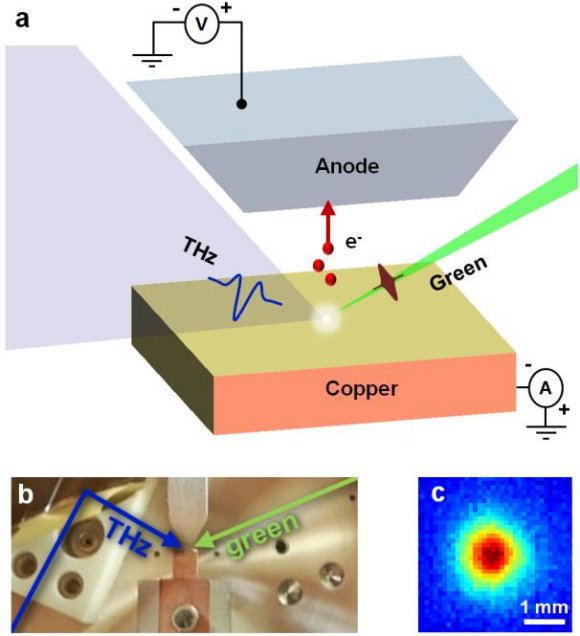

**Figure 1 | Experimental setup. (a) Electrons are emitted from a flat copper cathode by a 525 fs green (515 nm) pulse and accelerated by a p-polarized single-cycle terahertz pulse. Electron spectrum information is obtained by applying a retarding bias and measuring the photocurrent. (b) Photograph of apparatus inside a vacuum chamber. (c) Terahertz beam intensity profile at the focus.**

To provide basic understanding of the electron dynamics before describing experimental results, we present in Fig. 2 a simulation of emission and acceleration by a SC THz field for our experimental conditions. Figure 2(a) shows the THz field at the photocathode surface as a function of time. The electron momentum gain $p$ is described by the THz electric vector potential:

$p = -eA = -e \int_{t_{emit}}^{t} E_{THz}(t')dt'$, and the kinetic energy gain at nonrelativistic energies is described by

$KE = m_0 c^2 \sqrt{1 + \left(\frac{p}{m_0 c}\right)^2} - m_0 c^2 \cong \frac{p^2}{2m_0}$. The time-energy evolution of a single electron exposed to the

THz field is shown in Fig. 2(b) for two field strengths. Photoemission occurs at -0.47 ps, the first node of the THz electric waveform where the field switches from positive to negative (green dots in Fig. 2(a)-(c)). Shortly after emission, for a THz field of 72 MV/m (blue curve), the electron is accelerated to 120 eV by the main negative half-cycle of the THz waveform (-0.47 to 0.47 ps) and thereafter decelerated to 26 eV by the final positive half-cycle (0.47 – 1.43 ps). Figure 2(c) shows the time evolution of the energy spectrum for an electron bunch of 50 fC. Due to the long 525 fs green pulse (over which the electrons are emitted), some electrons are emitted during a phase of the THz which does not maximize the vector potential and therefore receive less acceleration. As a result, the spectrum has a mean energy below that given by the vector potential, unlike [33]. The evolution of the electron bunch is determined in two stages. At short time scales (<3 ps), the strong THz field dominates the energy of the bunch and the bunch behaves largely as that of the single electron. At long time scales (>3 ps, after the THz has passed), space charge forces cause leading/trailing electrons to gain/lose energy, resulting in a chirped energy spectrum. We obtain a mean electron energy of 20 eV and a peak energy of 90 eV (Fig. 2(d)). Given that the energy gain scales quadratically with the accelerating field in the nonrelativistic regime, the red curve in Fig. 2(b) and plots in Figs. 2(e) and 2(f) show that a 2 GV/m field will achieve a mean energy of 27 keV. Additionally, because the electrons are quickly accelerated out of the low-energy regime, the RMS energy spread is reduced to 3.5%. By modifying the anode and using a shorter green pulse duration, a mean energy of 99 keV and spread of 1.3% can be achieved (see Methods "2 GV/m THz gun").

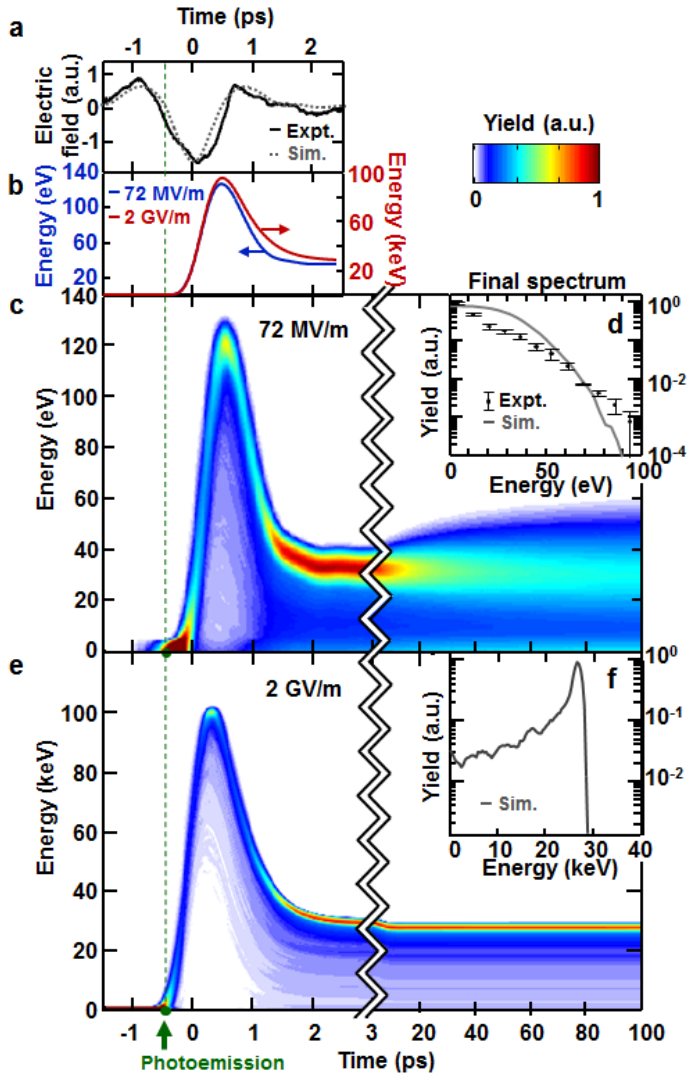

**Figure 2 | Electron acceleration in a single-cycle terahertz field.** (a) Electro-optic sampling trace of terahertz pulse. (b) Single-electron energy evolution in time. After emission (indicated by the green dashed line), the electron is accelerated by the negative half-cycle of the THz electric field and thereafter decelerated by the final positive half-cycle. Energies gained from a 72 MV/m (blue) and 2 GV/m (red) field are 26 eV and 27 keV, respectively. (c) 50 fC electron bunch energy evolution under a 72 MV/m THz field showing the effect of finite electron bunch duration and space charge. At long time scales, space charge induces an energy chirp on the spectrum. (d) The final energy spectrum (gray line, simulation; black dots with one sigma error, experiment) shows an mean energy of 20 eV (18 eV in experiment) with a peak energy of 90 eV (92 eV in experiment).

**Experimental data discussed further in Fig. 3(a). (e) 50 fC electron bunch energy evolution under a 2 GV/m THz field demonstrating the vanishing of the space charge-induced energy spread. (f) Final energy spectrum (simulation) showing an electron bunch at 27 keV with 3.5% RMS energy spread.**

The performance of the THz gun was determined by the electron energy spectra of the emitted electron bunch. The spectra were measured by taking the derivative of the photocurrent as a function of bias voltage. At reverse biases, the bias acts as a highpass filter for the photocurrent and therefore allows the measurement of the yield of electrons above an energy equal to the reverse bias times the unit charge. This setup essentially functions like a retarding field analyzer without a grid [34].

In Fig. 3(a), the experimentally measured spectrogram is plotted as a function of energy and delay (between the green pulse and THz pulse) for 0, 36, and 72 MV/m THz field strengths. Note, a positive delay corresponds to the situation where the THz pulse arrives after the green pulse and vice versa. At a delay of 0.25 ps at 72 MV/m, we observe photocurrent up to 92 eV while the mean energy was measured to be 18 eV. This is in good agreement with the simulations shown in Fig. 2(c) that predict a peak energy of 90 eV and a mean energy of 20 eV.

The pulsewidth of emission is 17% of the THz carrier period and is therefore capable of revealing THz electric field phase-dependent effects. Figure 3(b) plots the photocurrent as a function of delay at a fixed bias voltage of -0.1 V. A similarity between the photocurrent trace and the integrated THz field is observed. Given the fact that the photocurrent is proportional to the mean electron energy in this bias regime, the data reflects that the electron momentum gain is governed by the vector potential, $p = -e \int_{t_{emit}}^{\infty} E(t')dt'$.

In Fig. 3(c), we compare the bias sweep with THz OFF and THz ON (36 MV/m) at a delay of 0.25 ps. We observe a significant rise of photocurrent at reverse biases up to ~-30 V in the THz ON case, indicating clear evidence of THz acceleration. In Fig. 3(d), the energy spectra is in good agreement with the main features (mean/peak energy and spectral shape) of the simulation results. Finally, the increase in total photocurrent at Bias>0 for THz ON in Fig. 3(c) is due to the strong THz field acting in conjuction with the static field to increase the space-charge limited charge density [35], allowing more electrons to escape the cathode.

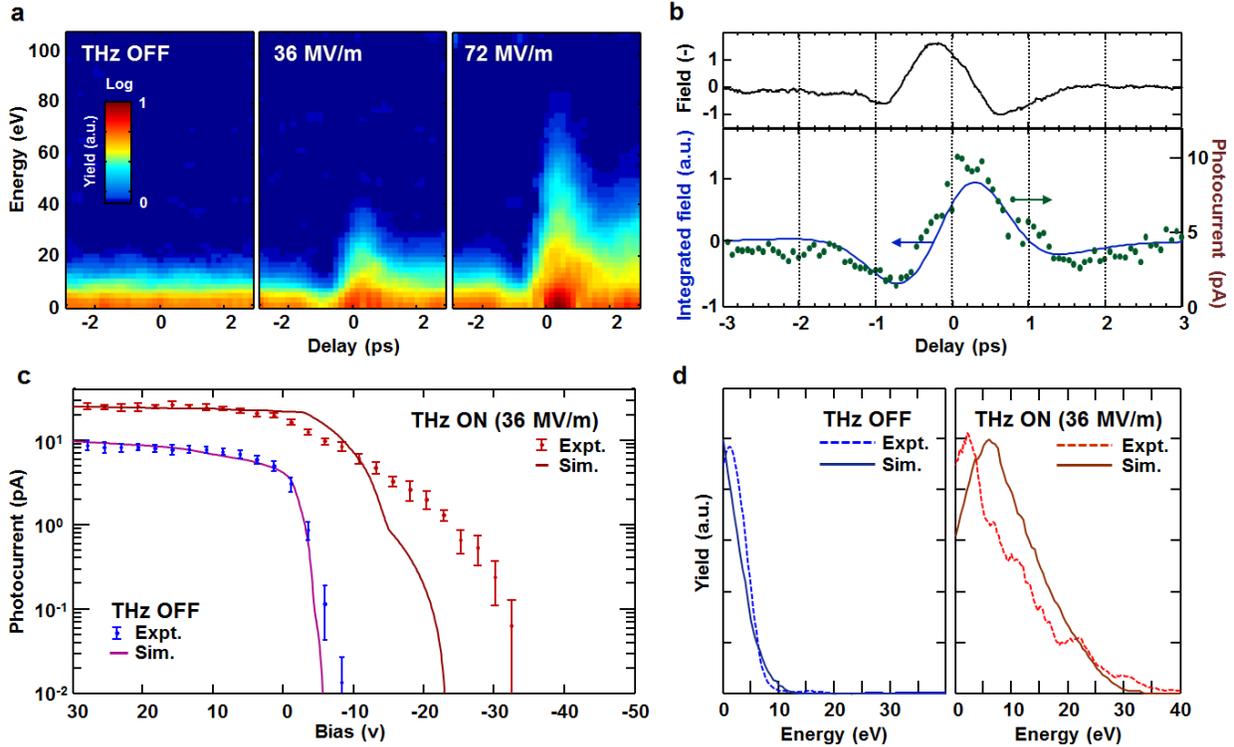

**Figure 3 | Experimental evidence of terahertz-driven electron acceleration. (a) Spectrograms showing electron acceleration at various THz field intensities. At 72 MV/m, we observe an increase of photocurrent at reverse biases up to -92 V, indicating that electrons achieved a peak energy of 92 eV to overcome the potential barrier. (b) Experimental correlation between the photocurrent and integrated terahertz field reflects that the acceleration is governed by the vector potential. The bias here is set to a regime where the photocurrent is proportional to the electron energy. (c) Bias sweeps of photocurrent (dots with one sigma error bar) show a significant rise in photocurrent at strong reverse biases indicating clear evidence of THz acceleration. (d) Electron energy spectra comparing THz OFF and THz ON (36 MV/m).**

In conclusion we have constructed a first version of a compact THz-driven electron gun. From a simple flat copper photocathode, we demonstrated a proof-of-concept THz gun accelerating a 50 fC electron bunch in agreement with particle tracking simulations. While the proof-of-concept does not yet compare to the performance of state-of-the-art DC guns [36] or RF guns [37], THz guns holds promise due to its orders-of-magnitude higher field breakdown threshold [22,38]. Further, we showed that with upgrades to readily-available GV/m THz sources and shorter photoemission laser pulses (20 fs), this scheme can yield monoenergetic electrons with energies up to 100 keV, comparable to the performance of current guns used for ultrafast electron diffraction. In combination with recent demonstration of THz waveguide linacs, the presented work also paves the way for the development of an all-optical, ultracompact, high repetition rate keV-MeV beamline.

**Methods**

**Terahertz beam**. The THz pulses are generated by optical rectification (OR) of tilted-pulse-front NIR pulses in cryogenic lithium niobate (LN). Up to 12 µJ THz pulse energy is generated with a conversion efficiency of about 1.0% at the crystal. The diverging THz beam is collimated by a 50.8 mm high-density polyethylene lens, converted from s- to p-polarization by a periscope, coupled into the vacuum chamber, and focused onto the photocathode by a parabolic mirror with an effective focal length of 25.4 mm. The THz beam is incident onto the photocathode at an angle of 67° (largest angle by which a f/1 focused beam can be incident on a flat surface without clipping). The large angle of incidence maximizes the electric field component normal to the photocathode surface.

The THz pulse is temporally characterized by a standard electro-optic (EO) sampling setup employing a 200 µm thick, 110-cut ZnTe crystal. 70 fs NIR pulses from the mode-locked fiber oscillator (see Pump source) sample the THz field-induced birefringence as a function of delay. A quarter-wave plate followed by a polarizer converts the field-induced birefringence to an intensity modulation, and the intensity modulation is recorded by a photodiode. The measured pulse has a carrier frequency of 0.45 THz and a FWHM bandwidth of 0.4 THz. The THz beam was characterized spatially by a pyroelectric detector array (Spiricon Pyrocam III).

**Pump source.** The pump source for THz generation is a Yb:KYW chirped pulse regenerative amplifier (RGA) producing 1.5 mJ pulses with 1 kHz repetition rate at a near-infrared (NIR) center wavelength of 1030 nm and bandwidth of 2.1 nm. The dielectric grating compressor following the RGA compresses the pulses to a transform-limited pulse duration of 750 fs. The seed for the RGA was a mode-locked Yb-doped fiber oscillator emitting 70 fs, 0.2 nJ pulses at 80 MHz amplified to 1.6 nJ by a Yb-doped fiber amplifier [39]. After losses through the optical elements in the setup, the impinging pump energy into the (LN) crystal was 1.2 mJ.

**Photoemitter.** About 2% of the available NIR pump energy was used to generate the green photoemitter pulses by second harmonic generation in a 0.75 cm long BBO crystal. A BG-39 bandpass filter was used to remove the fundamental NIR component. The green pulses have a FWHM duration of 525 fs and are focused at an angle of 67° onto the photocathode with a beam waist of 25 μm. Since the emitted charge scales as the green pulse's intensity squared due to two-photon ionization, the effective duration of the electron bunch at emission is estimated to be lower than the duration of the green pulse approximately by a factor of $\sqrt{2}$, corresponding to 375 fs. Since both the THz and green pulses are produced from the same 750 fs NIR laser, the timing jitter between them is negligible.

**Anode and cathode.** A Poisson equation solver (Superfish [40]) was used to model the static DC electric potential between the anode and cathode and it confirmed that the bias field is uniform over the emission area. Varying the DC bias between the cathode and anode does not impact the electron dynamics during the exposure to the THz field for two reasons. First, the bias was swept over +/- 110 V, corresponding to a DC field of +/- 52 kV/m, which is more than three orders of magnitude weaker than the THz field. Second, during the transient interaction with the THz pulse on the picosecond timescale, the electron dynamics are dominantly governed by the THz field.

**Simulations.** A particle tracking simulation was implemented to model the evolution of an electron bunch in the presence of the THz field. The initial electron bunch had a Gaussian spatial profile with beamwaist of 50 µm and a Gaussian temporal profile with FWHM of 375 fs. The initial kinetic energy was 0.18 eV (green two-photon energy subtracted by the copper work function) and the initial velocity vectors were uniformly distributed over a hemisphere. The particles were released 0.1 nm away from the surface. 6000 macroparticles were used to represent a total bunch charge of 50 fC, corresponding to a charge of 52e$^-$ per macroparticle, where e$^-$ is the unit charge. The trajectories were modeled by integrating the kinematic equations for every particle $i$ using a 4$^{th}$ order Runge-Kutta solver.

$$m\frac{d\boldsymbol{v}_i}{dt} = \boldsymbol{F}_{field} + \boldsymbol{F}_{bias} + \boldsymbol{F}_{image,i} + \sum_j \boldsymbol{F}_{ij}$$

$$\frac{d\boldsymbol{r}_i}{dt} = \boldsymbol{v}_i$$

Here, $m$ is the relativistic mass, $\boldsymbol{F}_{field}$ is the electric force due to the THz pulse, $\boldsymbol{F}_{bias}$ is the electric force due to the DC bias, $\boldsymbol{F}_{image,i}$ is the image charge force on the $i^{th}$ particle due to the bulk metal cathode, and $\boldsymbol{F}_{ij}$ is the particle-particle Coulomb force. The THz beam was modeled as a fundamental Gaussian beam with numerical aperture of 1/2. The THz pulse spectrum was directly adapted from that obtained by EO sampling with a flat spectral phase profile.

**2 GV/m THz gun.** With increased accelerating field strength, one can consider a way to prevent the electrons from being decelerated by the final positive half-cycle of the THz pulse. If the anode-cathode spacing is reduced to the distance that the electrons travel over the accelerating half-cycle (83 µm for a 2 GV/m field), then the electrons would be recombined on the anode before the onset of the decelerating half-cycle. Equivalently, a small sub-THz wavelength hole can be drilled into the anode to allow the electrons to escape the THz field region and be used for downstream applications, as illustrated in Fig. 4(b).

To reduce the energy spread, one can reduce the duration of the photoemitter pulse, such that the electrons are subject to the same vector potential. Figure 4(a) shows the evolution over distance of an electron bunch with an initial duration of 20 fs and transverse size of 12 μm. An energy of 99 keV is achieved over a 83 μm distance. At 83 μm (gray dashed line), the bunch enters a hole in the anode and escapes the decelerating THz half-cycle. The RMS spread at 125 μm is 1.3%, as shown in Fig. 4(c).

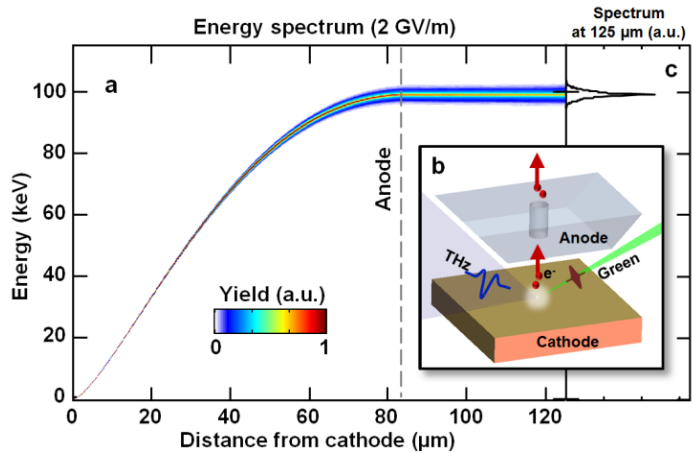

**Figure 4 | 2 GV/m THz gun. (a) Electron bunch spectrogram as a function of distance from cathode in the presence of a 2 GV/m THz field. The bunch is accelerated by the THz negative half-cycle to 99 keV over 83 μm. At 83 μm (gray dashed line), it enters a hole in the anode and escapes the decelerating half-cycle. (b) A schematic of the modified anode with a hole. (c) Energy spectrum at 125 μm from cathode. The RMS energy spread is 1.3%. For this simulation, a green pulse duration of 20 fs and spot size of 12 μm were used. Other parameters such as the THz waveform and bunch charge of 50 fC were preserved.**

**Acknowledgements**

We acknowledge Michael Swanwick and Peter R. Krogen for experimental assistance, Eduardo Granados for initial contributions to the project, and William S. Graves and Erich P. Ippen for helpful discussions. This work was supported by DARPA under contract N66001-11-1-4192, by the Air Force Office of Scientific Research under grant AFOSR - A9550-12-1-0499, the European Research Council through Synergy Grant 609920, and the excellence cluster "The Hamburg Centre for Ultrafast Imaging- Structure, Dynamics and Control of Matter at the Atomic Scale" of the Deutsche Forschungsgemeinschaft. W.R.H.

acknowledges support from the National Defense Science and Engineering Fellowship. L. J. W. acknowledges support from the Agency for Science, Technology and Research (A*STAR), Singapore.

**Author contributions**

W.R.H., K-H.H., and F.X.K. conceived and designed the experiment. W.R.H. built and performed the experiment with help from E.A.N. and P.D.K. K.R. provided insights to optimize the terahertz source. E.A.N., K.R., K-H.H., P.D.K., and F.X.K. provided feedback to improve the experiment. W.R.H. developed and performed the simulations with help from E.A.N. and L.J.W. L.J.W. and A.F. performed supporting simulations. W.R.H. and E.A.N. analyzed the data and interpreted the results. W.R.H. wrote the manuscript with contributions from E.A.N. and K.R. and revisions by all. K-H.H., L.Z. and F.X.K. provided management and oversight to the project.

**Competing financial interests**

The authors declare no competing financial interests.

**Materials & Correspondence**

Correspondence and requests for materials should be addressed to F.X.K.